\documentclass[12pt]{iopart}
\usepackage{graphicx}
\begin{document}

\title[Theoretical analyses predict A20 regulates period of NF-$\kappa$B oscillation]{Theoretical analyses predict A20 regulates period of NF-$\kappa$B oscillation}

\author{Benedicte Mengel, Sandeep Krishna, Mogens H. Jensen and Ala Trusina*}

\address{Center for Models of Life,
Niels Bohr Institute, Blegdamsvej 17, 2100 Copenhagen, Denmark.}
\ead{*trusina@nbi.dk}

\begin{abstract}
The nuclear-cytoplasmic shuttling of NF-$\kappa$B is characterized by damped
oscillations of the nuclear concentration with a time period of
around 1-2 hours. The NF-$\kappa$B network contains several feedback loops
modulating the overall response of NF-$\kappa$B activity.
While I$\kappa$B$\alpha$ is known to drive and I$\kappa$B$\varepsilon$ is known to dampen the oscillations, the precise role of A20 negative feedback
remains to be elucidated. Here we propose a model of the NF-$\kappa$B system focusing on three negative feedback loops~(I$\kappa$B$\alpha$, I$\kappa$B$\varepsilon$ and A20) which capture the experimentally observed responses in wild-type and knockout cells. We find that A20, like I$\kappa$B$\varepsilon$, efficiently dampens the oscillations albeit through a distinct mechanism. In addition, however, we have discovered a new functional role of A20 by which it controls the oscillation period of nuclear NF-$\kappa$B. The design based on three nested feedback loops allows independent control of period and amplitude decay in the oscillatory response. Based on these results we predict that adjusting the expression level of A20, e.g. by siRNA, the period can be changed by up to a factor 2.

\end{abstract}

\vspace{2pc}
\maketitle

\section*{Introduction}

Nuclear Factor-kappa B, NF-$\kappa$B, is a family of dimeric transcription factors involved in a number of important processes such
as immune response, cellular growth and apoptosis~\cite{Pahl}.
NF-$\kappa$B regulates the expression of more than a hundred genes and is implicated in a large number of diseases, including cancer, heart diseases and asthma~\cite{Pahl}.
Nuclear translocation of NF-$\kappa$B, necessary for its activity, is triggered by a wide variety of stress signals: endotoxin LPS, cytokines IL-1 and the tumor necrosis factor (TNF).
Fluorescence imaging of the TNF-triggered NF-$\kappa$B activity in single mammalian cells
shows distinct "spiky" but asynchronous oscillations in the level of nuclear NF-$\kappa$B \cite{Nelson04}; populations of mouse fibroblast cells continuously exposed to TNF
exhibit damped and smooth -- probably due to population averaging -- oscillations in the nuclear NF-$\kappa$B concentration.
The production of damped oscillations with a time period of around 1.5 hours thus seems to be a robust characteristic of the NF-$\kappa$B system.
\begin{figure}[ht]
\begin{center}
\includegraphics[width=1.0\hsize]{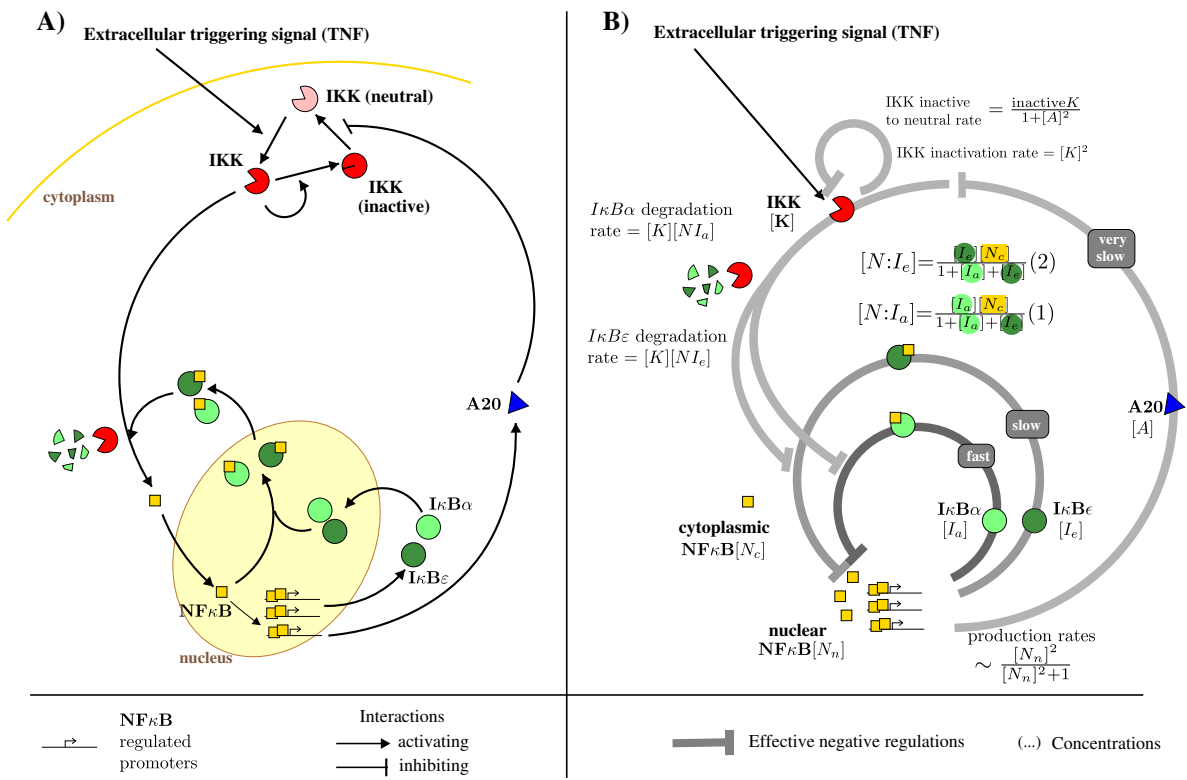}
\caption{Schematic drawing of the core of NF-$\kappa$B regulatory network. A) Details of the NF-$\kappa$B regulation. TNF activates the I$\kappa$B kinase (IKK) which in turn causes the phosphorylation, and subsequent degradation, of the I$\kappa$B inhibitor proteins, thus releasing NF-$\kappa$B. Free NF-$\kappa$B translocates to the nucleus inducing transcription of the inhibitor proteins,  I$\kappa$B$\varepsilon$ and I$\kappa$B$\alpha$, and A20. The I$\kappa$B proteins inhibit the NF-$\kappa$B transcription factor by actively exporting it out of the nucleus. A20 acts upstream by inactivating IKK. The I$\kappa$B$\beta$ is not shown.  B) Nested negative feedback  perspective on the NF-$\kappa$B regulation. NF-$\kappa$B is regulated by two parallel negative feedbacks acting through the inhibitor proteins I$\kappa$B$\alpha,\varepsilon$ and A20 negative feedback, acting upstream by inactivating IKK. The variation in the greyscale of the three feedback loops indicates the difference in timescales: dark grey stands for fast and light grey for slowest.} \label{fig.network}
\end{center}
\end{figure}
NF-$\kappa$B is regulated by several negative feedback loops: two acting through the inhibitor proteins I$\kappa$B$\alpha,\varepsilon$ which bind and sequester it in the cytoplasm. Another feedback regulates concentrations of nuclear NF-$\kappa$B through A20, see Fig.~\ref{fig.network}.
Addition of TNF activates the I$\kappa$B kinase (IKK) which in turn causes the phosphorylation, and subsequent degradation, of the I$\kappa$B inhibitor proteins, thus releasing NF-$\kappa$B. Free NF-$\kappa$B translocates to the nucleus inducing transcription of the inhibitor proteins,  I$\kappa$B$\varepsilon$ and I$\kappa$B$\alpha$, and A20. In turn the I$\kappa$B proteins inhibit the NF-$\kappa$B transcription factor by actively exporting it out of the nucleus. A20 on the other hand acts upstream by inactivating IKK, see Fig.~\ref{fig.network}A. There is another inhibitor protein, I$\kappa$B$\beta$,  but it is only slightly induced by NF-$\kappa$B compared to the two other inhibitor proteins~\cite{Kearns07} resulting in a weak negative feedback on NF-$\kappa$B. We will thus omit it in our model, see Fig.~\ref{fig.network}B.

\begin{figure}[ht]
\begin{center}
\includegraphics[width=0.8\hsize]{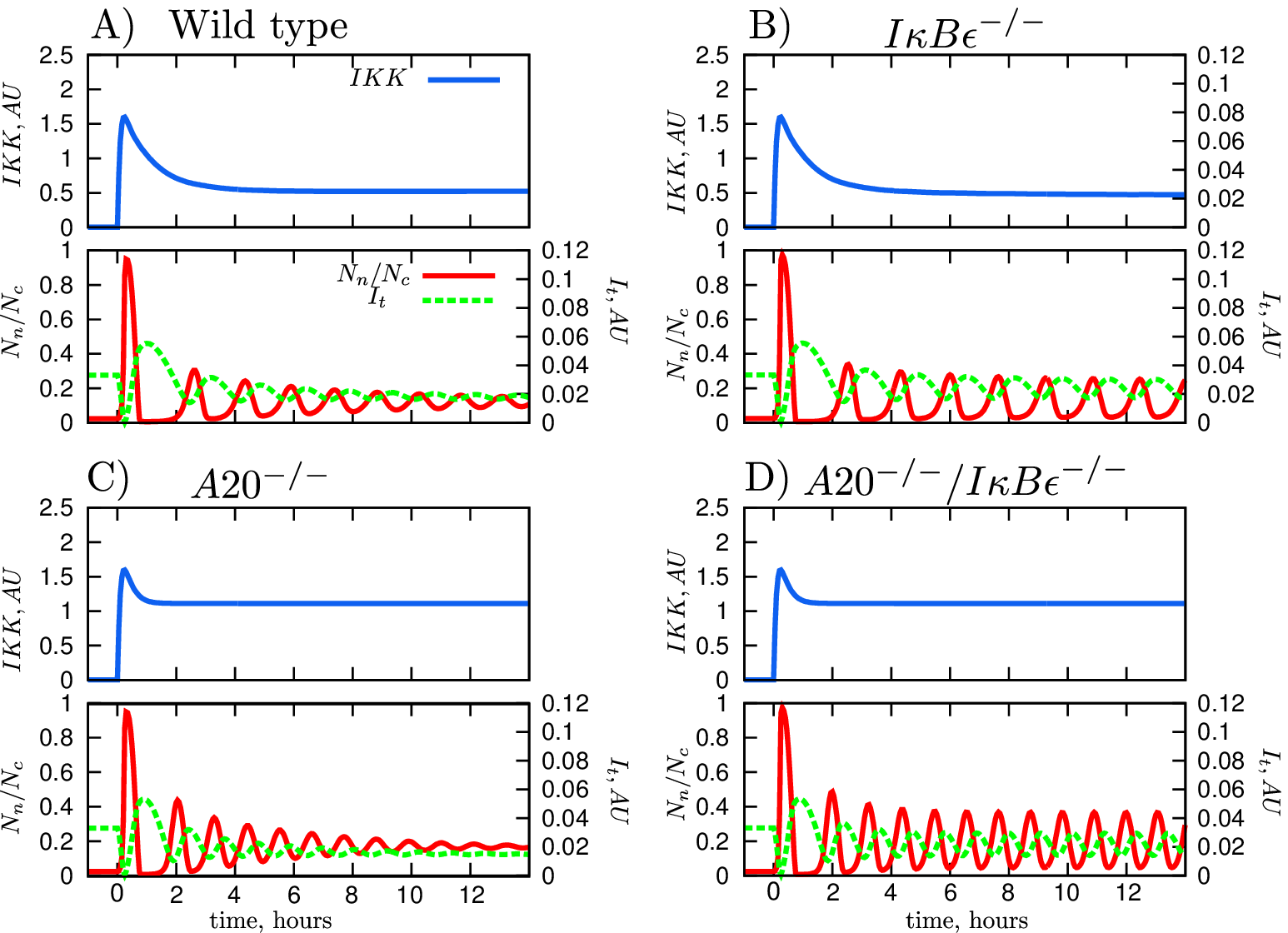}
\caption{Model simulations of the four states of the cell obtained by numerical integration of the Ordinary Differential Equations describing the model in Methods section. A: wild-type, B: $I\kappa B\varepsilon^{-/-}$, C: $A20^{-/-}$, D: $A20^{-/-} / I\kappa B\varepsilon^{-/-}$. In each panel blue line is the concentration of active IKK, red is the normalized concentration of the nuclear NF-$\kappa$B and dashed green line is the concentration of total inhibitor proteins, $I_t$ = I$\kappa$B$\alpha$ + I$\kappa$B$\varepsilon$.
Note the differences in NF-$\kappa$B oscillation period and decrease in the amplitude in absence of A20 and $I\kappa B\varepsilon^{-/-}$ and increased levels of IKK in late stage for A20 knockout cells.
}\label{fig.response}
\end{center}
\end{figure}

\begin{figure}[ht]
\begin{center}
\includegraphics[width=0.4\hsize]{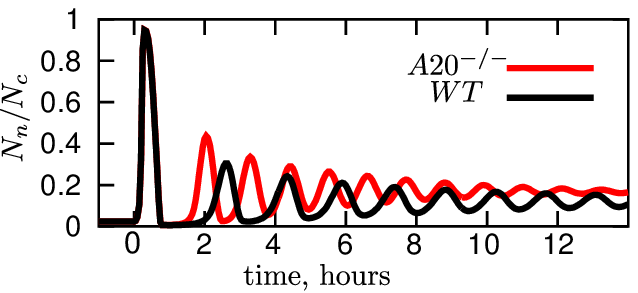}
\caption{A20 changes the period of nuclear NF-$\kappa$B oscillations. WT is in black, reproduced from Figure 2A, and A20 knockout is in red, reproduced from Figure 2C. Note the difference in steady states after oscillations are damped.}\label{fig.response_overlap}
\end{center}
\end{figure}

Each I$\kappa$B protein forms a negative feedback loop as they are all transcriptionally activated by NF-$\kappa$B.
The loops are not identical. Knockout mutant studies in bulk indicate that in the absence of I$\kappa$B$\varepsilon$ the nuclear NF-$\kappa$B oscillations are enhanced whereas there are no oscillations in the absence of I$\kappa$B$\alpha$. It has been suggested that I$\kappa$B$\varepsilon$ dampens oscillations generated by I$\kappa$B$\alpha$, which is consistent with the differences in their half-lives -- I$\kappa$B$\varepsilon$ is at least twice as stable as I$\kappa$B$\alpha$ \cite{Werner08}.
I$\kappa$B$\alpha$ is activated almost instantly while I$\kappa$B$\varepsilon$ activation occurs 37 min after NF-$\kappa$B enters the nucleus. This difference in half life and time of activation allows us to define I$\kappa$B$\alpha$ as a "fast" and I$\kappa$B$\varepsilon$ as a "slow" negative feedback, see Fig.~\ref{fig.network}.

A20 feedback acts upstream of NF-$\kappa$B and I$\kappa$B. It is an important regulator of late IKK activity and was shown experimentally to be required for the drop in NF-$\kappa$B activity separating early and late phase response to TNF when measured in bulk~\cite{Werner08}. Cells deficient in A20 show persistent IKK activity and develop severe inflammation and cachexia~\cite{Lipniacki04}.

The physiological importance of NF-$\kappa$B transcription factor and its intriguing dynamical behavior made it a center of attention for decades both from an experimental and theoretical point of view~\cite{Baltimore93,Ghosh98}. The first computational model of the NF-$\kappa$B pathway was proposed in Hoffmann et. al~\cite{Hoffmann02} and used to understand the dynamical responses of the NF-$\kappa$B wild-type and I$\kappa$B knockout, e.g. oscillations and their absence in knockouts. This model has later been modified and used by Nelson et al~\cite{Nelson04} to analyze oscillations in single cells. Krishna et al 2006~\cite{Krishna06} showed that the model can be significantly reduced while still capturing the essential dynamical features, in particular showing spiky oscillations in single cells.

Both modeling and experimental results suggest that A20 is important for lowering the level of nuclear NF-$\kappa$B after TNF stimulus~\cite{Lipniacki04, Werner08}. Other studies have focused on details of where and how A20 acts in the pathway~\cite{Ashall09}. To address the discrepancy between bulk and single cell data Ref.~\cite{Kim09, Ashall09} introduced stochasticity and showed that averaging single cell stochastic dynamics leads to a smooth damped response in bulk.

In the current view of the system, I$\kappa$B$\alpha$ feedback drives the oscillations and the I$\kappa$B$\varepsilon$ feedback dampens the oscillations. From the bulk experiments A20 is known to lower the level of nuclear NF-$\kappa$B after stimulus. It is not known how the temporal profile of NF-$\kappa$B in single cells is affected by A20~\cite{Werner08}. We would like to investigate whether and how A20 is modifying NF-$\kappa$B oscillatory behavior in single cells. To address this question we will extend the NF-$\kappa$B model by Krishna et al.~\cite{Krishna06} to include I$\kappa$B$\varepsilon$ and A20 negative feedbacks.

\section*{Model}

Figure~\ref{fig.network} shows schematic representation of the model.
It contains three negative feedback loops centered around NF-$\kappa$B: I$\kappa$B$\alpha$, I$\kappa$B$\varepsilon$ and A20.

The dynamical variable of most importance is $N_n$, the nuclear NF-$\kappa$B concentration.
The first term in the equation for $N_n$ is the rate of increase in nuclear NF-$\kappa$B concentration
due to import of free NF-$\kappa$B from the cytoplasm. This rate is lower for higher levels of the I$\kappa$B
proteins. The other two negative terms model the decrease of the nuclear concentration due
to sequestration by the I$\kappa$Bs and subsequent export into the cytoplasm. Over the timescales we
are interested in there is no significant production or degradation of NF-$\kappa$B.
The mRNA levels of I$\kappa$B$\alpha$ and I$\kappa$B$\varepsilon$ are regulated through a sigmoidal function of NF-$\kappa$B, given by $\frac{N_n^2}{N_n^2 + K^2}$.
Here we assumed that there is a weak cooperativity in NF-$\kappa$B activating transcription of I$\kappa$B$\alpha$/$\varepsilon$ with Hill coefficient two.
We also include a small basal level of transcription in I$\kappa$B$\alpha$, which has little influence on NF-$\kappa$B dynamics but has an important role in
reproducing the I$\kappa$B$\alpha$ and I$\kappa$B$\varepsilon$  mRNA fold induction experimentally measured in \cite{Kearns07}.

At the protein level, the rate of protein {\emph increase} is linearly proportional to the respective mRNA.
The rate of decrease in I$\kappa$B$\alpha$ is controlled by IKK-independent degradation, $I_a/\tau_\alpha$ and IKK-dependent degradation. The rate of I$\kappa$B$\alpha$ decay is proportional to both IKK and the concentration of complexes formed between I$\kappa$B$\alpha$ and cytoplasmic NF-$\kappa$B, $N_c$, $\propto IKK [I\kappa B\alpha:N_c]$.
Assuming that reaction rates for complex formation are much faster then the nuclear import/export and I$\kappa$B$\alpha$ degradation, the concentration of $[I\kappa B\alpha$:$N_c]$ can be derived to be
$(1-N_n)/(1 + I\kappa B\varepsilon + I\kappa B\alpha )$ (see Supplementary Materials for more details).
The changes in the I$\kappa$B$\varepsilon$ protein levels are governed by the same terms as for I$\kappa$B$\alpha$, but with different rate constants, which overall make the negative feedback through I$\kappa$B$\varepsilon$ slower and weaker (see Supplementary Material for details). Apart from this, the equations for I$\kappa$B$\varepsilon$ are the same as for I$\kappa$B$\alpha$.

\begin{figure}[h]
\begin{center}
\includegraphics[width=0.8\hsize]{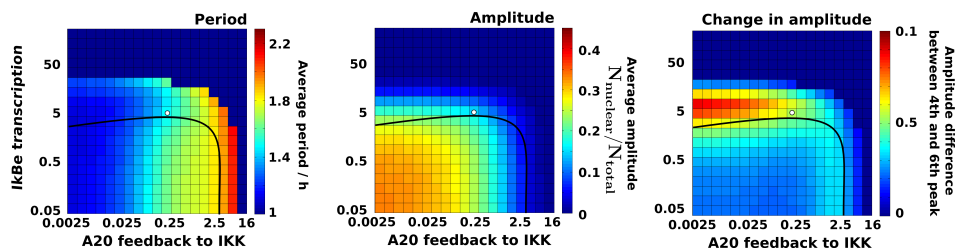}
\caption{Analysis of the  $I\kappa B\varepsilon^{-/-}$ and A20 feedback loops.
The strength of each loop is varied by varying the I$\kappa$B$\varepsilon$ transcription and the A20 coupling to IKK. The white circle represents the reference state of wild type (Figure 2A). The colorcoded are: A) The average period taken between the 2nd and 3rd peak. B) The amplitude of NF-$\kappa$B averaged over 2-nd and 3rd peak with a lower boundary of 1\%. C) The dampening effect, tells how fast are oscillations damped,  measured by  difference in amplitude between the 2nd and 4th peak. Bifurcation line is shown in black and separates regions of sustained oscillations from regions of damped oscillations. This  bifurcation line has been found by counting the number of peaks in the nuclear concentration of NF-$\kappa$B within the simulated timespan (see Methods section for details). Both I$\kappa$B$\varepsilon$ transcription and A20 coupling to IKK is able to change the amplitude~(B) and dampen the oscillations~(C) but only the A20 coupling to IKK is able to alter the period of the oscillations~(A).}\label{fig.heatmap}
\end{center}
\end{figure}

The I$\kappa$B Kinase~(IKK) is the driving force of the system as its activation leads to the degradation of the inhibitor proteins, the I$\kappa$Bs, and thereby the release of NF-$\kappa$B. IKK is activated upon stimulation of the membrane receptor but the detailed mechanism for this activation remains to be clarified. We have chosen to use the mechanism earlier proposed in Ref.~\cite{Ashall09} to model the IKK activation: a three step process where IKK is converted from its neutral state to being active by the triggering signal, TNF, see Fig.~\ref{fig.network}A. The active I$\kappa$B kinase can turn itself off and go back into neutral state before being activated again by the TNF signal. The I$\kappa$B kinase is shut down by A20 which inhibits the transformation from inactive to neutral IKK thereby leaving IKK in an inactive state, see Fig.~\ref{fig.network}A.

The first 30 minutes of IKK adaption-like temporal profile in response to TNF stimulation --
which appears to be independent of A20 regulation -- is modeled such that IKK peaks after about 15 minutes
and then goes to a new steady state after 30 minutes of TNF induction.
The new steady state is however determined by A20:
it is high in the absence of A20 and decreases with increasing concentrations of A20.
This is a slow feedback as IKK must first activate NF-$\kappa$B leading to the production of A20 that in turn shuts down the pathway.

We have taken most rates and timescales from existing literature wherever possible
and manually adjusted so that the model reproduces the following experimental observations:
\begin{itemize}
\item [1.] Wild-type cells show damped oscillations in nuclear NF-$\kappa$B with a time period of 60-100 min.
\item[2.] Mutants with I$\kappa$B$\alpha$ alone show enhanced oscillations with same period of 60-100 min, while those with I$\kappa$B$\varepsilon$ alone do not show oscillations.
\item[3.] The fold induction in  mRNA of I$\kappa$B$\varepsilon$ reaches twice the level of fold induction of the I$\kappa$B$\alpha$ mRNA
\end{itemize}

\section*{Results and discussion}
\subsection*{Model validation}
The basic response of our model, with the default parameters, to a continuous presence of TNF
is damped oscillations of nuclear NF-$\kappa$B. The original wild type response as well as I$\kappa$B$\varepsilon$ and I$\kappa$B$\alpha$ knockout
matches the experimental observations (see Fig.~\ref{fig.response}A, B), Supplementary Fig.~S1 and interactive applet\cite{applet}.
The time period of the oscillations is about 100 minutes, which also matches
experiments. Thus, the basic response (criterion 1 and 2) is correctly reproduced by the model, Fig.~\ref{fig.response}.
We also checked the model against criteria 3 and other known knockouts, for further details see Supplementary material.

\subsection*{A20 changes the period of nuclear NF-$\kappa$B oscillations}

\begin{figure}[h]
\begin{center}
\includegraphics[width=0.7\hsize]{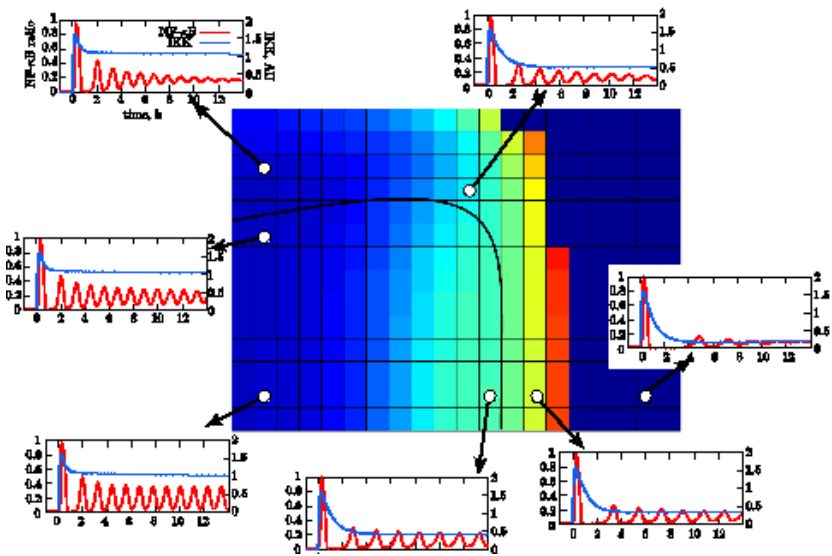}
\caption{Set of characteristic responses in nuclear NF-$\kappa$B obtained at different I$\kappa$B$\varepsilon$ transcription rates and A20 coupling strengths to IKK. The colorcoded is the average period taken between the 2nd and 3rd peak~(a section of A in Fig.~\ref{fig.heatmap}). Several parameter sets have been selected both within and outside of the bifurcation border~(the black line). Both I$\kappa$B$\varepsilon$ transcription and A20 coupling to IKK is able to change the amplitude and dampening of the nuclear transcription factor activity but only the A20 coupling to IKK is able to alter the period of the oscillations.}\label{fig.heatmap_tc}
\end{center}
\end{figure}

It has been suggested that A20 lowers nuclear NF-$\kappa$B levels of the late phase of the response. This observation stems from bulk experiments~\cite{Werner05,Werner08}. However, the role of A20 has not been investigated in single cells where NF-$\kappa$B has oscillatory response~\cite{Ashall09}.
Using the proposed model we aimed to investigate the role of A20 as a modifier of the NF-$\kappa$B oscillatory behavior.
We find that A20 is able to adjust the period of nuclear NF-$\kappa$B oscillations in the range from 1 to 3 hours.
Below we describe the details of our finding.

We have modeled the NF-$\kappa$B response in A20 knockout cells and were surprised to find that not only is the resting level of the late phase of the NF-$\kappa$B response increased (as shown in Fig.~S2)
but also the oscillations become more pronounced when compared to wild-type cells,  see Fig.~\ref{fig.response} A,C and Fig.~\ref{fig.response_overlap}.
The amplitude of the NF-$\kappa$B response is decreased in the presence of A20 whereas the period is increased. This effect of A20 dampening the oscillations and increasing the period is even more visible in I$\kappa$B$\varepsilon$ knockout cells where the oscillations are sustained, compare Fig.~\ref{fig.response}B and D.

A20 only has an effect on the temporal profile of IKK in its late phase: the A20 mRNA level peaks at 30~min and the protein shows its effect after 45-60~min~\cite{Werner05,Werner08}. This feature is captured by the model where the level of IKK in the late phase is pushed down in the presence of A20 and generates low frequency NF-$\kappa$B oscillations. In the absence of A20 the late phase of IKK stays at a high level and generates high frequency oscillations, compare Fig.~\ref{fig.response}A and C and Fig.~\ref{fig.response_overlap}.

\subsection*{I$\kappa$B$\varepsilon$ does not change the period of NF-$\kappa$B oscillations}
Is A20 feedback loop dispensable and can I$\kappa$B$\varepsilon$ have similar effect on the oscillations period? There are two possible mechanisms for I$\kappa$B$\varepsilon$ to dampen I$\kappa$B$\alpha$ oscillations: a) destructive interference, where I$\kappa$B$\varepsilon$ oscillates on the same time scale as I$\kappa$B$\alpha$ but with the shifted phase compared to I$\kappa$B$\alpha$, such phase shift or delay has been observed experimentally\cite{Kearns07,Ashall09} and b) where I$\kappa$B$\varepsilon$ varies on a slower time scale and the interference between fast changing I$\kappa$B$\alpha$ and slow I$\kappa$B$\varepsilon$ results in damped oscillations. In order for a) to be true I$\kappa$B$\varepsilon$ and I$\kappa$B$\alpha$ must exhibit comparable strength and frequency and thus I$\kappa$B$\alpha$ knockout should oscillate in single cells, however at present there is no such data available and bulk experiments show no oscillations in the I$\kappa$B$\alpha$ knockout. On the other hand it is known that I$\kappa$B$\varepsilon$  has a slower degradation through IKK dependent mechanism \cite{Whiteside97} which makes I$\kappa$B$\varepsilon$ change on a slower time-scale. Furthermore the b) scenario can easily reproduce bulk data for I$\kappa$B$\alpha$ knockout and therefore we model I$\kappa$B$\varepsilon$ through scenario b). With  this constraint I$\kappa$B$\varepsilon$ does not influence the frequency in our model (compare \ref{fig.response}A and B).

\subsection*{Function of A20 and I$\kappa$B$\varepsilon$ is robust to variation in parameter values}
Our observation that A20 decreases the frequency of NF-$\kappa$B oscillations whereas I$\kappa$B$\varepsilon$ is not capable of the same effect,  is based on the model with specific set of parameters that were chosen to fit experimental results.
To check if our observation is not merely a result of a specific parameter combination  we have investigated how the period,  amplitude as well as the oscillation dampening
change with varying some of the model parameters. In Figure~\ref{fig.heatmap}A we show how the period changes as the strengths of A20 and I$\kappa$B$\varepsilon$ feedback loops vary 100 fold above and below the values we used in Figure~\ref{fig.response}. The period is almost independent of the strength of I$\kappa$B$\varepsilon$ whereas it gradually increases as the strength A20 feedback loop increases, thus supporting our conclusion that A20 is a key regulator of the frequency of NF-$\kappa$B oscillations.
It is important to note that although it is possible to decrease the frequency by increasing the strength of the I$\kappa$B$\varepsilon$ loop when I$\kappa$B$\varepsilon$ transcription rate is larger than 50 (see top left corner in  Figure~\ref{fig.heatmap}A), the resulting oscillations are strongly damped as illustrated in Figure~\ref{fig.heatmap}C.

\subsection*{Reasons for the differential effect of A20 and I$\kappa$B$\varepsilon$ on the period}
We can understand why A20 and not I$\kappa$B$\varepsilon$ feedback can effectively adjust the frequency by looking at the IKK dependent degradation of I$\kappa$B$\alpha$ and I$\kappa$B$\varepsilon$ terms:
$$\mathrm{I\kappa B\alpha\ \ degradation:} \ \  \sim [K]\frac{[N_c][I_a]}{1 + [I_a] + [I_e]}$$ and
 $$\mathrm{I\kappa B \varepsilon \ \ degradation:} \ \ \sim [K]\frac{[N_c][I_e]}{1 + [I_a] + [I_e]}$$

IKK acts similarly on both I$\kappa$B$\varepsilon$ and I$\kappa$B$\alpha$ feedbacks: Higher levels of IKK correspond to faster degradation of I$\kappa$B$\varepsilon/\alpha$ and thus shorter period. In case of I$\kappa$B$\varepsilon$, however, --  because of competitive binding to cytoplasmic NF-$\kappa$B -- higher levels of I$\kappa$B$\varepsilon$ will lead to faster degradation of I$\kappa$B$\varepsilon$ but slower degradation of I$\kappa$B$\alpha$, see equations above.
Thus I$\kappa$B$\varepsilon$ can only increase the period when oscillations are driven by I$\kappa$B$\alpha$, whereas IKK can work in both directions.
There are, however, limits to how much IKK can increase the period. This limit is set by the fact that as A20 increases, and IKK decreases,  the system passes through a Hopf bifurcation (shown by black line in Figure ~\ref{fig.heatmap}) where it goes from sustained to damped oscillations so that the amplitude and period of oscillations are strongly correlated: the longer the period the smaller the amplitude, compare Figure ~\ref{fig.heatmap}A and B.

\subsection*{Conclusions}

Growing evidence indicates that temporal control of NF-$\kappa$B and the downstream genes is of crucial importance for cell functioning: constitutively active NF-$\kappa$B
is a cause of many human tumors. Active NF-$\kappa$B turns on the expression of genes that keep the cell proliferating and protect the cell from conditions that would otherwise cause it to die via apoptosis. At the same time defects inactivating NF-$\kappa$B  result in increased susceptibility to apoptosis leading to increased cell death.
It appears that the original solution to this dilemma is through  transient activation of NF-$\kappa$B \cite{Hoffmann02} which on a single-cell level presents as damped oscillations.
Such temporal control can allow for selective gene activation \cite{Werner05, Ashall09}.
Given that the NF-$\kappa$B temporal response is of a high importance and is primarily regulated by three negative feedback loops we investigated the role of each negative feedback in shaping the response with the main focus on A20 negative feedback.

We find that the design of having two I$\kappa$B feedback loops on the same level as NF-$\kappa$B and the upstream feedback from A20 allows for a wide variety in possible NF-$\kappa$B outputs. The fast feedback from I$\kappa$B$\alpha$ generates the oscillatory behavior of the transcription factor which is damped by the delayed and out of phase activity of I$\kappa$B$\varepsilon$. This general response of NF-$\kappa$B upon TNF stimulation is altered by the upstream feedback from A20 which is slower than the two other feedbacks and acts on the I$\kappa$B kinase and not on the transcription factor itself. By changing the IKK profile, A20 is able to modulate the frequency of the transcription factor in a way not possible from the two I$\kappa$B feedbacks. Additionally, A20 lowers the end level of the nuclear NF-$\kappa$B meaning that when the system comes to rest more NF-$\kappa$B is removed from the nucleus, in the presence of A20, compare Fig.~\ref{fig.response} A and C and Fig.~\ref{fig.response_overlap} \cite{Werner08}. (This is because IKK is less active and thus not degrading the I$\kappa$Bs as fast, leaving the inhibitor proteins to enter the nucleus and export the transcription factor).

 If the biphasic response -- as seen in bulk experiments --  is a result of the population average of single cells with oscillating NF-kB, then, in bulk experiments, A20 will exhibit its effect by affecting the timing of the second phase onset.
 Thus a specific prediction would be that -- as the timing between first two peaks is shorter, see Supplementary Figure S3,  --  the second phase should start earlier in A20 knockout cells.

We found that similarly to I$\kappa$B$\varepsilon$,  A20 is able to dampen oscillations albeit through a different mechanism.
Whereas I$\kappa$B$\varepsilon$ dampens oscillations through competition with I$\kappa$B$\alpha$ for binding to NF-$\kappa$B and acts about an hour after TNF induction,  A20 dampens oscillations by inhibiting active IKK during late phase of the response, after about 2-3 hours.
A distinct novel finding of our investigation suggests that not only does A20 dampen the oscillations, it can also adjusts the period of the oscillations.
Thus, A20 together with I$\kappa$B$\varepsilon$ allows independent tuning of both dampening -- through I$\kappa$B$\varepsilon$ negative feedback --  and the frequency -- through A20 negative feedback.
This combination of nested feedback loops covers a wider variety of temporal responses where one can access both sustained oscillations and  damped oscillations with low or high frequency.

Our findings lead to a clear prediction that in single cells decreasing the coupling between A20 and IKK should lead to higher frequency oscillations in NF-$\kappa$B.
This can be experimentally tested by knocking down A20 with siRNA. Our model also predicts that the frequency of oscillations can be further increased by increasing the TNF dose in A20 knockout cells, where the IKK level in the late phase of the response is proportional to TNF dose.
An interesting future direction would be to examine how the diversity of NF-$\kappa$B oscillating temporal profiles created by nested feedback loops can allow for selective gene activation.

\ack
We thank Alexander Hoffmann for stimulating discussions. This work was funded by the Danish National Research Foundation.

\appendix

\section*{Methods}
\subsection*{Model Description}
\begin{eqnarray}
\frac{dN_n}{dt} = A\frac{(N_t-N_n)} {(K_I+ \eta I_{\alpha}+\eta I_{\varepsilon})} - B (I_{\alpha} + I_{\varepsilon})\frac{ N_n}{(\delta+N_n)}
\\
\frac{I_{m\alpha}}{dt} = p + t_a \frac{ N_n^2}{(K_D^2 + N_n^2)} - \gamma_{m\alpha}I_{m\alpha}\\
\frac{dI_{\alpha}}{dt} = k_{tla}I_{m\alpha} - \alpha_{\alpha}K\frac{(N_t-N_n)I_{\alpha}}{(K_I+\eta I_{\alpha}+\eta I_{\varepsilon})}- \gamma_\alpha I_{\alpha}\\
\frac{dI_{m\varepsilon}}{dt} = t_e \frac{N_n^2}{(K_D^2 + N_n^2)} - \gamma_{m\varepsilon}I_{m\varepsilon}\\
\frac{dI_{\varepsilon}}{dt} = k_{tle}I_{m\alpha} - \alpha_{\varepsilon}K\frac{(N_t-N_n)I_{\varepsilon}}{(K_I+\eta I_{\alpha}+\eta I_{\varepsilon})}-\gamma_\varepsilon I_{\varepsilon}\\
\frac{dA_m}{dt} = t_{A} N_n^2- \gamma_{Am}A_m\\
\frac{dA}{dt} = k_{tlA}A_m - \gamma_{A} A\\
\frac{dK}{dt} = T(K_t - K - K_i ) - \mu K^2\\
\frac{dK_i}{dt} =  \mu K^2 - \beta\frac{K_i}{(\sigma A^2 + 1)}\\
\label{eq.supmodel1}
\end{eqnarray}

The equations are all rescaled, see Supplementary Materials for details
of the rescaling and the parameter values.\\
\newline
\begin{table*}
\begin{center}
\begin{tabular}{|l|l|l|}
\hline
\textbf{Variable} &  \textbf{Description} & \tabularnewline 
\hline
\hline
$N_{n}$ & nuclear NF-$\kappa$B  &\tabularnewline 
\hline
$I_{\alpha/\varepsilon}$ & free I$\kappa$B  &\tabularnewline 
\hline
$I_{m\alpha/\varepsilon} $ &  I$\kappa$B mRNA & \tabularnewline
\hline
$A_m $ &A20 mRNA&\tabularnewline
\hline
 K & active IKK  & \tabularnewline
\hline
 $K_i$& inactive IKK & \tabularnewline
\hline
\hline
\textbf{Parameter} & \textbf{Description} & \textbf{Value}\\
\hline
\hline
B  &  proportionality factor of the export of nuclear NF-$\kappa$B & 102.6\tabularnewline 
\hline
A  &  proportionality factor of the import of NF-$\kappa$B & 0.004\tabularnewline 
\hline
$\eta$ & & 0.092 \tabularnewline 
\hline
$\delta$ & concentration at which half of the I$\kappa$B$\alpha/\varepsilon$ is bound  & 0.0414~$\mu M$\\  
& in complex with NF-$\kappa$B& \tabularnewline
\hline
p  & NF-$\kappa$B in-dependent transcription rate of I$\kappa$B$\alpha$ mRNA & 3.36$\cdot 10^{-5}~min^{-1}$\tabularnewline 
\hline
$t_{a}$ & NF-$\kappa$B dependent transcription rate of I$\kappa$B$\alpha$ mRNA  & 0.0042~$\mu M~min^{-1}$ \tabularnewline
 \hline
$t_{e}$ & NF-$\kappa$B dependent transcription rate of I$\kappa$B$\varepsilon$ mRNA  & 0.084~$\mu M~min^{-1}$\tabularnewline 
\hline
$t_A $ & NF-$\kappa$B dependent A20 transciption rate & 0.0168~$\mu M^{-1}min^{-1}$\tabularnewline
\hline
$tl_{a} $ & translation rate of I$\kappa$B$\alpha$ & 0.0672~$min^{-1}$\tabularnewline
 \hline
$tl_{e} $ & translation rate of I$\kappa$B$\varepsilon $ & 1.2$\cdot10^{-5}$~$min^{-1}$\tabularnewline
 \hline
$tl_{A} $ & translation rate of A20 & 0.3024~$min^{-1}$\tabularnewline
 \hline

 $\gamma_{Im\alpha}$& half-life of I$\kappa$B$\alpha$ mRNA &  0.0168~$min^{-1}$\tabularnewline
 \hline
$\gamma_{Im\varepsilon}$& half-life of I$\kappa$B$\varepsilon$ mRNA& 0.00168~$min^{-1}$\tabularnewline
 \hline
$\gamma_{I\alpha/\varepsilon}$& half-life of the I$\kappa$B's & 0.005~$min^{-1}$ \tabularnewline
 \hline
$\gamma_{A20m}$& half-life of the A20 mRNA& 0.00168~$min^{-1}$ \tabularnewline
 \hline
$\gamma_{A20}$& half-life of the A20& 0.00168~$min^{-1}$ \tabularnewline
 \hline
$\alpha_{\alpha} $ & IKK dependent degradation of I$\kappa$B$\alpha$ & 0.00025~$min^{-1}$\\
 \hline
$\alpha_{\varepsilon} $ & IKK dependent degradation of I$\kappa$B$\varepsilon$ & 0.03$\cdot$0.00025~$min^{-1}$\\
 \hline
$\mu$ & rate of IKK self-inactivation  & 0.063~$min^{-1}$\tabularnewline
\hline
$\sigma $ & strength of A20 negative feedback  & 0.25\tabularnewline
\hline

\end{tabular}
\caption{Model variables and parameters}
\end{center}
\end{table*}

\subsection*{Approximating Bifurcation Line}
To find the border for the bifurcation we counted the number of peaks in the nuclear concentration of NF-$\kappa$B. We counted the oscillations as sustained if the number of peaks was 100 or above. This approximation slightly overestimates the size of the region with sustained oscillations and is not affecting any results and conclusions.
\newpage
\section*{References}
\bibliographystyle{unsrt}

\bibliography{MSbib_2}
\newpage
\section*{Supplementary Materials}

\begin{figure}[ht]
\begin{center}
\includegraphics[width=0.9\hsize]{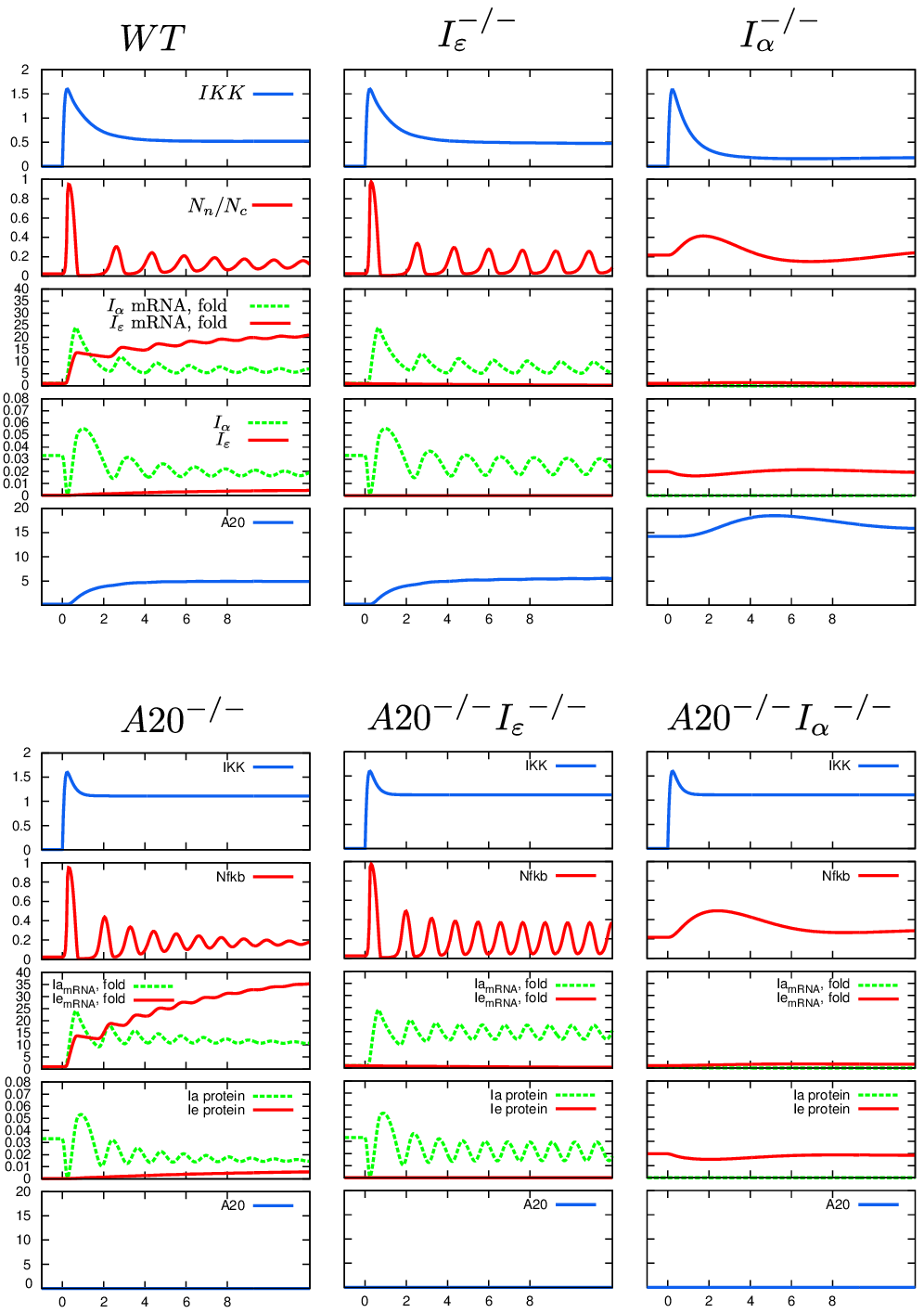}
\caption{Model simulations of the four states of the cell : A: wild-type, B: $I\kappa B\varepsilon^{-/-}$, C: $A20^{-/-}$, D: $A20^{-/-} / I\kappa B\varepsilon^{-/-}$}\label{fig.response_sup}
\end{center}
\end{figure}

\begin{figure}[ht]
\begin{center}
\includegraphics[width=0.9\hsize]{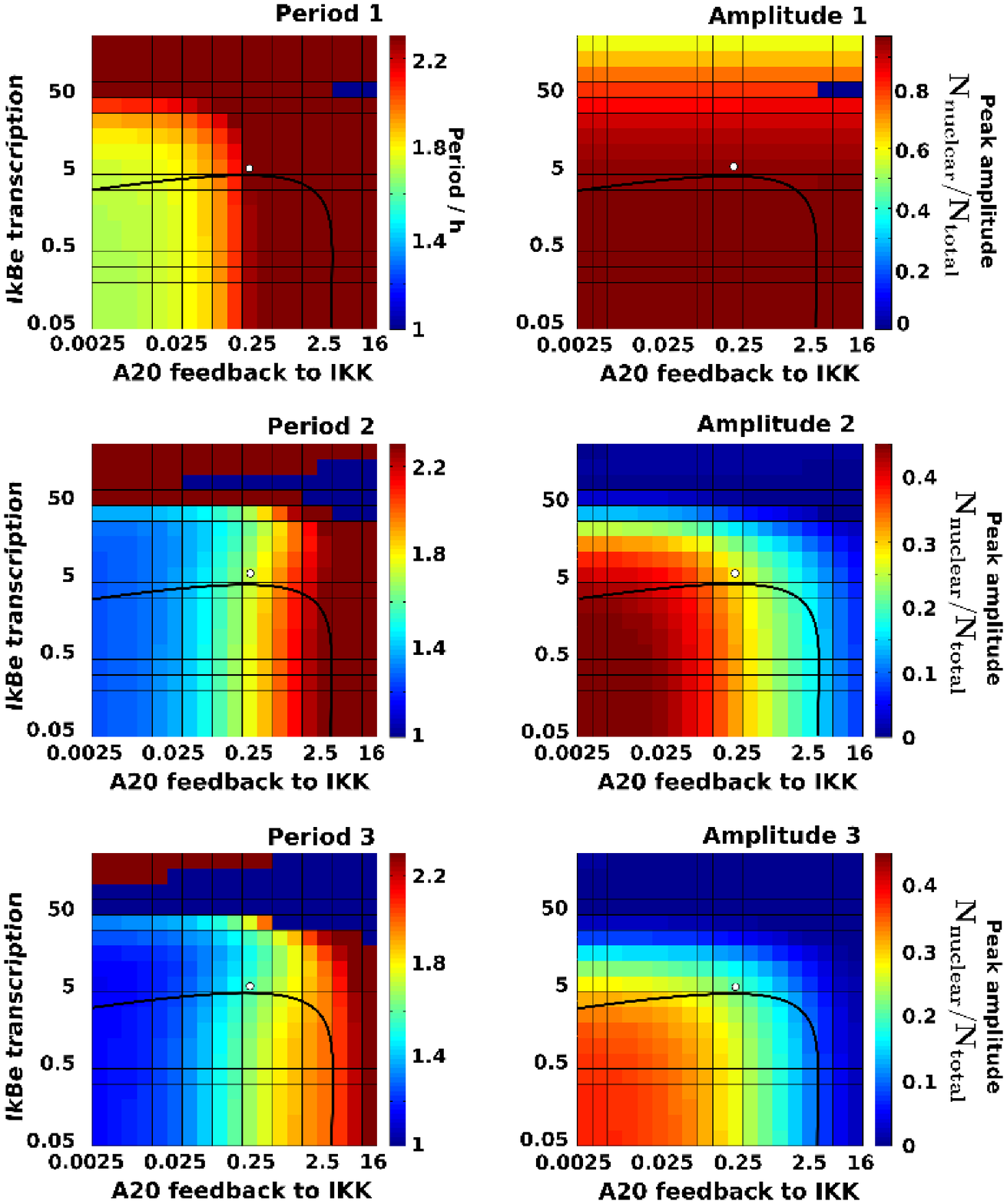}
\caption{Model simulating the wild-type cell. Left column: Recordings of the durations of the $1^{st}$, $2^{nd}$ and $3^{rd}$ period measures as the time between first and second peak, second and third and trird and fourth peak. Right column: Amplitude of $1^{st}$, $2^{nd}$ and $3^{rd}$peak. }\label{fig.response_sup}
\end{center}
\end{figure}
\newpage
\subsection*{Initial Model for I$\kappa$Bs regulation}
We use the following abbreviations:
$N_n \& N$, free nuclear and cytoplasmic
NF-$\kappa$B;
$I_m$, I$\kappa$B mRNA; $I_n \& I$, free nuclear and cytoplasmic I$\kappa$B;
$(NI)_n \& (NI)$, nuclear and cytoplasmic NF-$\kappa$B--I$\kappa$B complex;
IKK, I$\kappa$B kinase.

The equations for I$\kappa$B$\varepsilon$ are the same as for
I$\kappa$B$\alpha$ so we just use variable $I$ in our derivations.

The seven-variable model is defined by the equations
$$\frac{dN_n}{dt}=k_{Nin}N-k_{fn}N_nI_n+k_{bn}(NI)_n,$$
$$\frac{dI_m}{dt}=k_tN_n^2-\gamma_mI_m,$$
$$\frac{dI}{dt}=k_{tl}I_m-k_fNI+k_b(NI)-k_{Iin}I+k_{Iout}I_n,$$
$$\frac{dN}{dt}=-k_fNI+(k_b+\alpha)(NI)-k_{Nin}N,$$
$$\frac{d(NI)}{dt}=k_fNI-(k_b+\alpha)(NI)+k_{NIout}(NI)_n,$$
$$\frac{dI_n}{dt}=k_{Iin}I-k_{Iout}I_n-k_{fn}N_nI_n+k_{bn}(NI)_n,$$
$$\frac{d(NI)_n}{dt}=k_{fn}N_nI_n-(k_{bn}+k_{NIout})(NI)_n.$$


\subsection*{Reduced Model for I$\kappa$Bs regulation}

First, taking note of the fact that $k_f$ and $k_{fn}$ are large, we assume that
all complexes are in equilibrium, i.e.
$$k_fNI\approx(k_b+\alpha)(NI),$$
$$k_{fn}N_nI_n\approx(k_{bn}+k_{NIout})(NI)_n.$$
Simulations show that these are good approximations.
In terms of $I_n^{tot}\equiv I_n+(NI)_n$ and $N_c^{tot}\equiv N+(NI)=N_{tot}-N_n$, which are slowly varying,
we can rewrite the above equations as follows:
$$(NI)=(N_{tot}-N_n)\frac{I}{K_I+I},$$
$$N=(N_{tot}-N_n)\frac{K_I}{K_I+I},$$
$$(NI)_n=I_n^{tot}\frac{N_n}{K_N+N_n},$$
$$I_n=I_n^{tot}\frac{K_N}{K_N+N_n},$$
where $K_I\equiv(k_b+\alpha)/k_f=0.035~\mu M$ and $K_N\equiv(k_{bn}+k_{NIout})/k_{fn}=0.029~\mu M$, using the parameter values above.

Using these expressions, the equations of the seven-variable model reduce to the following four (Fig. 9):
$$\frac{dN_n}{dt}=k_{Nin}K_I\frac{(N_{tot}-N_n)}{K_I+I}-k_{NIout}\frac{I_n^{tot} N_n}{K_N+N_n},$$
$$\frac{dI_m}{dt}=k_tN_n^2-\gamma_mI_m,$$
$$\frac{dI}{dt}=k_{tl}I_m-\alpha \frac{(N_{tot}-N_n)I}{K_I+I}-k_{Iin}I+k_{Iout}K_N\frac{I_n^{tot}}{K_N+N_n},$$
$$\frac{dI_n^{tot}}{dt}=k_{Iin}I-k_{Iout}K_N\frac{I_n^{tot}}{K_N+N_n}-k_{NIout}\frac{I_n^{tot}N_n}{K_N+N_n}.$$

First, we note that the terms $-k_{Iin}I$ and $k_{Iout}K_N\frac{I_n^{tot}}{K_N+N_n}$
in the $dI/dt$ equation
are much smaller than $-\alpha \frac{(N_{tot}-N_n)I}{K_I+I}$ and can be neglected
as long as IKK is nonzero. Second, simulations reveal that the term
$k_{NIout}\frac{I_n^{tot} N_n}{K_N+N_n}$, in the $dI_n^{tot}/dt$ equation, also shows sharp spikes as a function of time
which coincide with the spikes of $N_n$. The value of this term is substantial
only when $N_n\gg K_N$, i.e., during the spikes of $N_n$, and at those times $I_n^{tot}$ dips to its minimum.
We therefore make the approximation that $I_n^{tot}$ can be
replaced by its minimum value, $I_{n,min}^{tot}$, which satisfies the equation
$$k_{Iin}I=k_{Iout}K_N\frac{I_{n,min}^{tot}}{K_N+N_n}+k_{NIout}\frac{I_{n,min}^{tot}N_n}{K_N+N_n}.$$
In the regime where $N_n\gg K_n$ this gives
$$I_{n,min}^{tot}\approx \frac{k_{Iin}}{k_{NIout}}I.$$

Using this we can reduce to a three-variable model
$$\frac{dN_n}{dt}=k_{Nin}K_I\frac{(N_{tot}-N_n)}{K_I+I}-k_{Iin}\frac{IN_n}{\delta+N_n},$$
$$\frac{dI_m}{dt}=k_tN_n^2-\gamma_mI_m,$$
$$\frac{dI}{dt}=k_{tl}I_m-\alpha\frac{(N_{tot}-N_n)I}{K_I+I}.$$

\subsection*{Parameters and variables re-scaling}
We start with the following system of equations

\begin{eqnarray}
\frac{dN_n}{dt} = k_{im}K_I\frac{(N_t-N_n)} {(K_I+ I_{\alpha}+I_{\varepsilon})} - B (I_{\alpha} + I_{\varepsilon})\frac{ N_n}{(\delta+N_n)}
\\
\frac{I_{m\alpha}}{dt} = p + t_a \frac{ N_n^2}{(K_D^2 + N_n^2)} - \gamma_{m\alpha}I_{m\alpha}\\
\frac{dI_{\alpha}}{dt} = k_{tla}I_{m\alpha} - \alpha_{\alpha}K\frac{(N_t-N_n)I_{\alpha}}{K_I+I_{\alpha}+I_{\varepsilon}}-\frac{I_{\alpha}}{\gamma_\alpha}\\
\frac{dI_{m\varepsilon}}{dt} = t_e \frac{N_n^2}{(K_D^2 + N_n^2)} - \frac{I_{m\varepsilon}}{\gamma_{m\varepsilon}}\\
\frac{dI_{\varepsilon}}{dt} = k_{tle}I_{m\alpha} - \alpha_{\varepsilon}K\frac{(N_t-N_n)I_{\varepsilon}}{K_I+I_{\alpha}+I_{\varepsilon}}-\frac{I_{\varepsilon}}{\gamma_\varepsilon}\\
\frac{dA_m}{dt} = t_{A} N_n^2/K_{DA}- \frac{A_m}{\gamma_{Am}}\\
\frac{dA}{dt} = k_{tlA}A_m - \frac{A}{\gamma_{A}}\\
\frac{dK}{dt} = T(K_t - K - K_i ) - \mu K^2\\
\frac{dK_i}{dt} =  \mu K^2 - \beta\frac{K_i}{\sigma A^2 + 1}\\
\label{eq.supmodel1}
\end{eqnarray}

where $k_{im}$ is the import of the NfkB into the nucleus, $K_I$ is dissociation constant for IkBs binding to Nfkb, ...

Re-define
\begin{itemize}
\item[1]
Re-define $N_n$ by scaling with $K_D$ and $N_t$
$$N_n'=N_n/K_D, \Rightarrow N_t' = N_t/K_D,  \delta' = \delta/K_D, \alpha_\alpha' = \alpha_\alpha/K_D,  \alpha_\varepsilon' = \alpha_\varepsilon/K_D$$  and furthermore,
$$N_n'' = N_n'/N_t' = N_n/N_t,  \delta'' = \delta'/N_t' = \delta/N_t$$
After this transformations and redefining $N_n''=N_n, \delta'' = \delta, etc. $the equations will be rescaled to :
$$\frac{dN_n}{dt} = k_{im}K_I\frac{(1-N_n)} {(K_I+ I_{\alpha}+I_{\varepsilon})} - B(I_{\alpha} + I_{\varepsilon})\frac{ N_n}{(\delta+N_n)}\\$$
$$\frac{I_{m\alpha}}{dt} = p + t_a \frac{ N_n^2}{(1 + N_n^2)} - \gamma_{m\alpha}I_{m\alpha}\\$$
$$\frac{dI_{\alpha}}{dt} = k_{tla}I_{m\alpha} - \alpha_{\alpha}K\frac{(1-N_n)I_{\alpha}}{K_I+I_{\alpha}+I_{\varepsilon}}-\frac{I_{\alpha}}{\gamma_\alpha}\\$$
the equations for $I_\varepsilon$ and $I_{m\varepsilon}$ scaled similarly and the rest remains as above.
\item[2]Re-define $I_\alpha$ and $I_\varepsilon$ by scaling with $K_I$.\\
$I_{\alpha/\varepsilon} ' = \frac{I_{\alpha/\varepsilon}}{K_I}  \Rightarrow B' = B K_I,\ \ k_{tla/e}' = k_{tla/e}'/K_I,   \alpha_{\alpha/\varepsilon}' = \alpha_{\alpha/\varepsilon}/K_I$
\item[3]Re-define $I_{\alpha m}$ and $I_{\varepsilon m}$ by scaling with $k_{tla}'$.\\
Re-defining
$I_{\alpha/\varepsilon m}' = k_{tla/e}'I_{\alpha/\epsilon m} = k_{tla/e}/K_I I_{\alpha/\epsilon m}, \Rightarrow p' = p k_{tla/e}/K_I, t_{a/e}' = t_{a/e} k_{tla/e}/K_I$
results in following equations for
$$\frac{dN_n}{dt} = k_{im}\frac{(1-N_n)} {(1+ I_{\alpha}+I_{\varepsilon})} - B(I_{\alpha} + I_{\varepsilon})\frac{ N_n}{(\delta+N_n)}\\$$
$$\frac{I_{m\alpha}}{dt} = p + t_a \frac{ N_n^2}{(1 + N_n^2)} - \gamma_{m\alpha}I_{m\alpha}\\$$
$$\frac{dI_{\alpha}}{dt} = I_{m\alpha} - \alpha_{\alpha}K\frac{(1-N_n)I_{\alpha}}{1+I_{\alpha}+I_{\varepsilon}}-\frac{I_{\alpha}}{\gamma_\alpha}\\$$
\item[4]Re-define $A_m$ by scaling with $k_{tlA}$.\\
$A_m' = k_{tlA}A_m, \Rightarrow  t_A' = k_{tla}t_A/K_{DA}, \sigma = \sigma/{k_{tla}}^2$ thus equations for A20 become
$$\frac{dA_m}{dt} = t_{A} N_n^2- \frac{A_m}{\gamma_{Am}}$$
$$\frac{dA}{dt} = A_m - \frac{A}{\gamma_{A}}$$
\item[5] Re-define $K$ and $K_i$ by scaling with $K_t$.\\
$K=K/K_t, \ \ K_i = K_i/K_t , \mu=K_t\mu$ thus equations for IKK become
$$\frac{dK}{dt} = T(1 - K - K_i ) - \mu K^2$$
$$\frac{dK_i}{dt} =  \mu K^2 - \beta\frac{K_i}{\sigma A^2 + 1}$$.

Thus the final system of equations is:
\begin{eqnarray}
\frac{dN_n}{dt} = k_{im}\frac{(1-N_n)} {(1+ I_{\alpha}+I_{\varepsilon})} - B(I_{\alpha} + I_{\varepsilon})\frac{ N_n}{(\delta+N_n)}\\
\frac{I_{m\alpha}}{dt} = p + t_a \frac{ N_n^2}{(1 + N_n^2)} - \gamma_{m\alpha}I_{m\alpha}\\
\frac{dI_{\alpha}}{dt} = I_{m\alpha} - \alpha_{\alpha}K\frac{(1-N_n)I_{\alpha}}{1+I_{\alpha}+I_{\varepsilon}}-\frac{I_{\alpha}}{\gamma_\alpha}\\
\frac{I_{m\varepsilon}}{dt} = t_e \frac{ N_n^2}{(1 + N_n^2)} - \gamma_{m\varepsilon}I_{m\varepsilon}\\
\frac{dI_{\varepsilon}}{dt} = I_{m\varepsilon} - \alpha_{\varepsilon}K\frac{(1-N_n)I_{\varepsilon}}{1+I_{\alpha}+I_{\varepsilon}}-\frac{I_{\varepsilon}}{\gamma_\varepsilon}\\
\frac{dA_m}{dt} = t_{A} N_n^2- \frac{A_m}{\gamma_{Am}}\\
\frac{dA}{dt} = A_m - \frac{A}{\gamma_{A}}\\
\frac{dK}{dt} = T(1 - K - K_i) - \mu K^2\\
\frac{dK_i}{dt} =  \mu K^2 - \beta\frac{K_i}{\sigma A^2 + 1}\\
\end{eqnarray}

and the scaled variables and parameters are:

$N_n = \frac{N_n}{N_t}$;
$I_{\alpha/\varepsilon} = \frac{I_{\alpha/\varepsilon}}{K_I}$;
$I_{m\alpha/\varepsilon} = \frac{k_{tla/e}}{K_I} I_{m\alpha/\varepsilon}$;
$A_m = k_{tlA}A_m$;
$K= \frac{K}{K_t}$;
$K_i= \frac{K_i}{K_t}$\\

$\delta = \frac{\delta}{N_t} = 0.0414$;
$B = BK_I = 0.014$;
$k_{tla/e} = \frac{k_{tla/e}}{K_I}; $;
$p = p\frac{k_{tla}}{K_I}  = 58.4$;
$t_{a/e} = t_{a/e}\frac{k_{tla/e}}{K_I}, t_a = 7300, t_e = 27.4 $;
$\alpha_{\alpha/\varepsilon}' = \frac{\alpha_{\alpha/\varepsilon}}{K_D K_I}, \alpha_\alpha = 219, \alpha_\varepsilon = 6.57 $;
$t_A = k_{tlA}\frac{t_A}{K_{DA}}  = 18$;
$\sigma = \frac{\sigma}{k_{tlA}^2} = 77x10^{-5} $;
$\beta = 5 $;
$\mu = K_t\mu = 100$;
TNF changes from 0.001 to 2.5;
$K_t = 6.67$\\
\end{itemize}

\begin{center}
\begin{tabular}{|l|l|}

\hline
\textbf{Variable} &  \textbf{Description} \tabularnewline
\hline
\hline
$N_{n}= \frac{N_n}{N_t}$ & nuclear NF-$\kappa$B normalized to total NF-$\kappa$B \tabularnewline
\hline
$I_{\alpha/\varepsilon} = \frac{I_{\alpha/\varepsilon}}{K_I}$ & free IkBs scaled with dissociation constant $K_I$ of I$\kappa$Bs binding to NF-$\kappa$B \tabularnewline
\hline
$I_{m\alpha/\varepsilon} = \frac{k_{tla/e}}{K_I} I_{m\alpha/\varepsilon}$ & re
-defined value of I$\kappa$B mRNA, $k_{tla/e}$ is the translation rate of I$\kappa$B$\alpha/\varepsilon$ \tabularnewline
\hline
$A_m = k_{tlA}A_m$ &re-defined A20 mRNA,  $k_{tlA}$ is the A20 translation rate\tabularnewline
\hline
 $K= \frac{K}{K_t}$& active IKK normalized to the total IKK, $K_t$ \tabularnewline
\hline
 $K_i= \frac{K_i}{K_t}$& inactive IKK normalized to the total IKK, $K_t$ \tabularnewline
\hline
\hline
\textbf{Scaled Parameter} & \textbf{Description}\\
\hline
$\delta = \frac{\delta}{N_t}$ ($\mu M^{-1}$)& concentration at which half of the I$\kappa$B$\alpha/\varepsilon$ is bound in complex with\\  & NF-$\kappa$B, normalized to total NF-$\kappa$B \tabularnewline
\hline
$B = BK_I$ &  proportionality factor of the export of nuclear NF-$\kappa$B, \\
 & scaled with the respective translation rates and dissoc. constant of\\
 & NF-$\kappa$B binding to IkBs, $K_I$ \tabularnewline 
\hline
$A = AK_I$ &  proportionality factor of the import of NF-$\kappa$B, \\
 & scaled with the respective translation rates and dissoc. constant of\\
 & NF-$\kappa$B binding to IkBs, $K_I$ \tabularnewline 
\hline
$p = p\frac{k_{tla/e}}{K_I} $ & constituitve, NF-$\kappa$B dependent transcription rate \\
&of IkBa mRNA, scaled with the respective translation rates and $K_I$ \tabularnewline
\hline
$t_{a/e} = t_{a/e}\frac{k_{tla/e}}{K_I} $ & NF-$\kappa$B dependent transcription rates of I$\kappa$Bs mRNA scaled with $K_I$\tabularnewline
  \hline
$t_A = k_{tlA}\frac{t_A}{K_{DA}} $ & A20 transciption rate scaled with A20 translation rate, $k_{tlA}$,\\
 &and diss. constant of NF-$\kappa$B binding to DNA at the operator site\\
 & controlling A20 promoter.  \tabularnewline
\hline
$\gamma_{Im\alpha}/\gamma_{Im\alpha}$& half-life of I$\kappa$B$\alpha$ mRNA scaled with $\gamma_{Im\alpha}$\tabularnewline
 \hline
$ \gamma_{Im\varepsilon}/\gamma_{Im\alpha}$& half-life of I$\kappa$B$\varepsilon$ mRNA scaled with $\gamma_{Im\alpha}$\tabularnewline
 \hline
$\gamma_{I\alpha/\varepsilon}\gamma_{Im\alpha}$& half-life of the I$\kappa$B's scaled with $\gamma_{Im\alpha}$\tabularnewline
 \hline
 $\gamma_{A20m}\gamma_{Im\alpha}$& half-life of the A20 mRNA scaled with $\gamma_{Im\alpha}$\tabularnewline
 \hline
$\gamma_{A20}\gamma_{Im\alpha}$& half-life of the A20 scaled with $\gamma_{Im\alpha}$\tabularnewline
 \hline
$\alpha_{\alpha/\varepsilon} = \frac{\alpha_{\alpha/\varepsilon}}{K_D K_I}$ & rate constant for IKK dependent degradation \\
&scaled with dissoc. constant of NF-$\kappa$B binding to \\
&operator site at the I$\kappa$B's promoter, $K_D$ and $K_I$  \tabularnewline
 \hline
$\mu = K_t\mu$ & rate of IKK self-inactivation scaled with total IKK, $K_t$ \tabularnewline
\hline
$\sigma = \frac{\sigma}{k_{tlA}^2} $ & strength of A20 negative feedback scaled \\
& with the square of A20 translation rate, $k_{tlA}^2$\tabularnewline
\hline
 \end{tabular}
\end{center}


%
 

\end{document}